\documentclass[astron]{eas}
\usepackage{hyperref}

\renewcommand{\[}{\begin{equation}}
\renewcommand{\]}{\end{equation}}
\def\p{\partial}

\let\boldgrk=\gkvecten
\let\boldgrksc=\gkvecseven

\def\gkthing#1{{\mathchoice%
	{\hbox{{\boldgrk\char#1}}}
	{\hbox{{\boldgrk\char#1}}}
	{\hbox{{\boldgrksc\char#1}}}
	{\hbox{{\boldgrksc\char#1}}}}}

\def\vtheta{\gkthing{18}}

{\newif\ifnotend
\notendtrue
\def\veclist{ABCDEFGHIJKLMNOPQRSTUVWXYZabcdefghijklmnopqrstuvwxyz.}
\def\top#1#2.{#1}
\def\tail#1#2.{#2.}
\loop\expandafter\xdef\csname v\expandafter\top\veclist\endcsname%
{{\noexpand\bf\expandafter\top\veclist}}
\edef\veclist{\expandafter\tail\veclist}
\if\veclist.\notendfalse\fi\ifnotend\repeat}
\def\d{{\rm d}}

\def\mnras{MNRAS}

\def\apj{ApJ}
\def\apjs{ApJS}

\def\bolOm{\mbox{\boldmath$\Omega$}}
\def\vOmega{\bolOm}

\def\kms{\,\mathrm{km\,s}^{-1}}

\def\Gevdens{\,\mathrm{GeV\,cm}^{-3}}

\def\d{\mathrm{d}}

\begin{document}
\title{Making action-angle disc models for Gaia} \author{Paul
  J. McMillan} \address{Lund Observatory, Lund University, Department
  of Astronomy and Theoretical Physics, Box 43, SE-22100, Lund,
  Sweden} 
\secondaddress{Rudolf Peierls Centre for Theoretical
  Physics, 1 Keble Road, Oxford, OX1 3NP, UK} 

\begin{abstract}
  I describe dynamical modelling of the Milky Way using action-angle
  coordinates. I explain what action-angle coordinates are, and what
  progress has been made in the past few years to ensuring they can be
  used in reasonably realistic Galactic potentials. I then describe
  recent modelling efforts, and progress they have made in
  constraining the potential of the Milky Way and the local dark
  matter density.
\end{abstract}
\maketitle

\section{Introduction}

When available, action-angle variables are the most convenient way of
describing orbits in a given gravitational potential $\Phi$.  The
three actions ($\vJ$) are constants of motion, and therefore label an
orbit. They are the conjugate momenta of the angle coordinates
$\vtheta$. Since $\vJ$ is constant, basic Hamiltonian mechanics allows
us to write that
\[
-{\p H \over \p \vtheta} = \dot{\vJ} = 0.
\]
This means that $H$ must be independent of $\vtheta$. The derivatives
of $H$ are therefore also clearly independent of $\vtheta$, so we can
again use basic Hamiltonian mechanics to write
\[
\dot{\vtheta} = {\p H \over \p \vJ} = \vOmega(\vJ) 
\]
where $\vOmega$ is known as the frequencies \emph{and is independent
  of $\vtheta$}. Therefore the angles of a particle on a given orbit
(i.e. fixed $\vJ$) increase linearly with time -- we can write this in
component form as
\[
 \theta_i(t) = \theta_i(0) + \Omega_i(\vJ) t.
\]
Since the orbit is bound and regular, the position and velocity of an
object is a \emph{periodic} function of the angles. This periodicity is
defined to be over $2\pi$ (so if we increase any angle coordinate by
$2\pi$ we return to the original position). 

The components of the actions can defined by
\[
  J_i  =  {1 \over 2\pi} \oint_{\gamma_i} \vJ \cdot \d\vtheta
  =   {1 \over 2\pi} \oint_{\gamma_i} \vp \cdot \d\vq 
\]
where the path $\gamma_i$ is one on which $\theta_i$ increases by
$2\pi$. (An expanation of why the second equality is true, along with
a much more detailed discussion of action-angle coordinates can be
found in Binney \& Tremaine \cite{GDII}.) For example, a path in an axisymmetric
potential that goes from $\phi=0$ to $\phi=2\pi$ (while other
coordinates are unchanged) is clearly one such path, we then have
\[
J_\phi = {1 \over 2\pi} \int_0^{2\pi} p_\phi\,\d\phi = {1 \over 2\pi}
\int_0^{2\pi} L_z\,\d\phi = L_z ,
\]
where $L_z$ is the angular momentum about the $z$ axis.

So, in a given potential, the values $\vJ$ label an orbit, and
$\vtheta$ labels a point on that orbit. 

The relationship between action-angle coordinates and the position and
velocity of a particle will clearly depend on the gravitational
potential in which the particle is moving. The major reason that
action-angle coordinates are not widely used in astrophysics is that
-- for most potentials -- it is not possible to determine them by simple
techniques. 

In a spherically symmetric potential two of the actions are simply
related to the total angular momentum and the angular momentum about
some preferred axis. The third action and the angles can be found by a
1D numerical integral. In the special case of the isochrone potential
these integrals can be performed analytically. 

This is possible because the equations of motion are separable in
spherical polar coordinates in a spherically symmetric potential. The
other instance where the equations of motion are separable is in the
St{\"a}ckel family of potentials  (e.g. de Zeeuw \cite{deZ85}). Again, in this
case, the angles and actions can be found by simple 1D integrals -- in
this case in confocal ellipsoidal coordinates.

\subsection{Finding action-angle coordinates for more realistic
  galactic potentials}
Over the past few years there has been substantial improvement in the
methods available for calculating action-angle coordinates in other
potentials. In the past is was common to estimate them using the
``adiabatic approximation'' (e.g. Binney \cite{JJB10}), which is the
approximation that the motion perpendicular to the Galactic plane can
be decoupled from the motion parallel to the plane. The ``St{\"a}ckel
fudge'' improves on this by approximating that the motion can be
decoupled in the ellipsoidal coordinates associated with St{\"a}ckel
potentials (Binney \cite{JJB12:Stackel}). This has now been extended to
triaxial potentials (Sanders \& Binney \cite{SaJJB15:Triaxial}).

Other methods require the use of so-called ``generating functions'' to
manipulate the known action-angle coordinates in a given potential
(typically the isochrone potential) so that they are valid in a new
potential. The ``torus method'' (e.g. McMillan, Binney et al, in prep
-- \url{https://github.com/PaulMcMillan-Astro/Torus}) uses this to determine
the full phase-space structure of an orbit with a given $\vJ$ in a
given potential. A new method based on an orbit integration (Sanders
\& Binney \cite{SaJJB14} -- \url{https://github.com/jlsanders/genfunc})
allows one to do similar given an initial position and velocity,
rather than a value $\vJ$.

These methods are now publicly available. The axisymmetric
``St{\"a}ckel fudge'' will be made available by Binney in the near
future. A version of the St{\"a}ckel fudge, and routines similar to
Sanders and Binney's orbit integration method are also available
through \textsc{galpy} (Bovy \cite{Bo15} --
\url{http://github.com/jobovy/galpy}). 

\section{Modelling}
Distribution functions (DFs) that are a function of action alone are
in equilibrium (this follows from Jeans' theorem).
Simple functional forms for the DF of a disc galaxy in equilibrium
have been in use for around 5 years (Binney \cite{JJB10}, though the commonly
used form is the altered version used by Binney \& McMillan \cite{JJBPJM11}). They are of
the form
\[
f(\vJ) = f_\phi(J_\phi)\,f_r(J_r,J_\phi)\,f_z(J_z,J_\phi)
\]
where $f_\phi$ primarily controls the radial surface density, $f_z$
primarily controls the vertical density and velocity profile, and
$f_r$ primarily controls the radial and azimuthal velocity
distributions. We take 
\[
f_r \propto \exp(-\kappa\,J_r/\sigma^2_r) ;\qquad 
f_z\propto \exp(-\nu\,J_r/\sigma^2_z).
\]
These have had substantial success in fitting the local velocity
distribution and density profiles (e.g. Binney \cite{JJB12:dfs}).
Since they ensure consistency between the radial and azimuthal
velocity distributions, initial attempts to fit the local velocity
distribution were unsuccessful. This was shown to be because the
peculiar velocity of the Sun differed by around $7\kms$ from the value
that was accepted at the time, and assumed in the initial analysis
(Binney \cite{JJB10}).

Alternative DFs for halo-like components have also been proposed and
used (Binney \cite{JJB14}, Posti et~al. \cite{Poea15}). The disc DFs
have been adapted to include velocity substructure (McMillan
\cite{PJM11:Hyades,PJM13:Hyades}), or information about chemistry
(Sanders \& Binney \cite{SaJJB15:EDF}).

The biggest reason to use DFs of the form $f(\vJ)$ is that it allows
one to learn about the gravitation potential that the stars are moving
in. To learn anything about the potential one has to start from the
approximation that the Galaxy is in equilibrium, as otherwise any set
of observed stellar positions and velocities are consistent with any
potential. 

It is not a simple task to fit observational data about the
Milky Way to these models, but a method was demonstrated on
mock data (McMillan \& Binney \cite{PJMJJB13}, see also Ting et~al. \cite{YSTea13}). The key point
recognised was that an approach based upon an orbit library was doomed
to fail, as the orbit library could never hope to be sufficiently
dense that each observed star had a reasonable number of orbits
sampling its error ellipse. The next step
was to do this for real data.

\section{The potential of the Milky Way from RAVE data}
A new study (Piffl et~al. \cite{Piea14}) uses data from the RAVE survey to
constrain the Milky Way's potential and the local dark matter
density. It does this by demanding that the derived DF and
gravitational potential satisfy 3 demands: 1) It must fit binned
velocity histograms taken from the RAVE survey (after allowing for
uncertainties). 2) The stellar density as a function of $z$ above the
sun due to the DF must match that of the stellar component of the mass
model that produces the potential. 3) This vertical density profile is
also fit to a vertical density profile for the Milky Way found from
the SDSS (Juri{\'c} et~al. \cite{Juea08_short}).  

This work was able to provide constraints on the Milky Way potential,
which are described in detail in the paper. The headline result is
that it found the local dark matter density (assuming an oblate or
spherical halo with axis ratio $q$) is
\[
\rho_{{\rm dm},\odot} = (0.48\times q^{-\alpha})\, \Gevdens
\]
with a systematic uncertainty of $15$ per cent. The main contribution
to the uncertainty is the uncertain distances to stars observed by
RAVE. This is an uncertainty that Gaia will dramatically reduce. 

It should also be noted that a similar study by Bovy \& Rix
(\cite{BoRi13}), using data from the Segue survey, found results
consistent with those found by Piffl et al.

\section*{Acknowledgements}
This work is a discussion and review of work I've undertaken in recent
years under the guidence of James Binney, and in collaboration with Til
Piffl, Jason Sanders and others in the Oxford dynamics group. This
research was supported by a grant from the Science and Technology
Funding Council.

\bibliographystyle{astron}

\begin{thebibliography}{}


\bibitem[2010]{JJB10} Binney, J., 2010, \mnras, 401, 2318

\bibitem[2012a]{JJB12:dfs}  Binney, J., 2012a,  \mnras, 426, 1328

\bibitem[2012b]{JJB12:Stackel} Binney, J.,  2012b,  \mnras,  426, 1324

\bibitem[2014]{JJB14} Binney, J., 2014,  \mnras,  440, 787

\bibitem[2011]{JJBPJM11} Binney, J. , McMillan, P., 2011, \mnras,  413, 1889

\bibitem[2008]{GDII} Binney, J. , Tremaine, S., 2008, {\em Galactic Dynamics, Second Edition,} Princeton University Press

\bibitem[2015]{Bo15} Bovy, J., 2015, \apjs , 216, 29

\bibitem[2013]{BoRi13} Bovy, J. , Rix, H.-W., 2013, \apj,  779, 115

\bibitem[1985]{deZ85} de Zeeuw, T., 1985, \mnras,  216, 273

\bibitem[2008]{Juea08_short} Juri\'c, M et al, 2008,  \apj,  673, 864

\bibitem[2011]{PJM11:Hyades} McMillan, P.~J., 2011,  \mnras,  418, 1565

\bibitem[2013]{PJM13:Hyades} McMillan, P.~J., 2013,  \mnras,  430, 3276

\bibitem[2013]{PJMJJB13} McMillan, P.~J. , Binney, J.~J., 2013, \mnras,  433, 1411

\bibitem[2014] {Piea14} Piffl, T., Binney, J., McMillan, P.~J., The RAVE collaboration, 2014, \mnras,  445, 3133

\bibitem[2015] {Poea15}
Posti, L. , Binney, J. , Nipoti, C. , Ciotti, L., 2015,
  \mnras,  447, 3060

\bibitem[2014]{SaJJB14} Sanders, J.~L. , Binney, J., 2014, \mnras,  441, 3284

\bibitem[2015a]{SaJJB15:Triaxial} Sanders, J.~L. , Binney, J., 2015a, \mnras,  447, 2479

\bibitem[2015b] {SaJJB15:EDF} Sanders, J.~L. , Binney, J., 2015b,  ArXiv e-prints

\bibitem[2013] {YSTea13} Ting, Y.-S., Rix, H.-W., Bovy, J., van de Ven, G., 2013, \mnras,  434, 652

\end{thebibliography}

\end{document}